\documentclass[prb,aps,amsmath,amssymb,nofootinbib,epsfig,showpacs,twocolumn,superscriptaddress]{revtex4-1}

\usepackage{amsmath,cleveref,amssymb,amsfonts,graphicx,multirow,color,bm,textcomp,mathtools}
\usepackage{paralist}
\usepackage{srcltx}
\usepackage[all,cmtip]{xy}
\usepackage{mathrsfs}
\usepackage{srcltx,soul,subfigure}
\usepackage{threeparttable,dsfont,bbm}

\crefname{equation}{Eq.}{Eqs.}
\crefname{figure}{Fig.}{Figs.}

\newcommand{\bra}[1]{\langle #1 |}
\newcommand{\ket}[1]{| #1 \rangle}

\newcommand{\bee}{\begin{eqnarray}}
\newcommand{\ee}{\end{eqnarray}}
\newcommand{\bma}{\begin{pmatrix}}
\newcommand{\ema}{\end{pmatrix}}
\newcommand{\balig}{\begin{align}}
\newcommand{\ealig}{\end{align}}

\newcommand{\ba}{\begin{align}}
\newcommand{\ea}{\end{align}}

\newcommand{\ignore}[1]{}

\usepackage{dcolumn}
\newcolumntype{C}[1]{>{\centering\let\newline\\\arraybackslash\hspace{0pt}}m{#1}}

\begin{document}

\title{Conductance interference  in a superconducting Coulomb blockaded Majorana ring}

\author{Ching-Kai Chiu}
\affiliation{
Condensed Matter Theory Center and Joint Quantum Institute and Station Q Maryland, Department of Physics, University of Maryland, College Park, MD 20742, USA}

\author{Jay D. Sau}
\affiliation{
Condensed Matter Theory Center and Joint Quantum Institute and Station Q Maryland, Department of Physics, University of Maryland, College Park, MD 20742, USA}

\author{S. Das Sarma}
\affiliation{
Condensed Matter Theory Center and Joint Quantum Institute and Station Q Maryland, Department of Physics, University of Maryland, College Park, MD 20742, USA}

\begin{abstract} 
	By tuning the magnetic flux, the two ends of a 1D topological superconductor weakly coupled to a normal metal as a ring-shaped junction can host split Majorana zero modes (MZMs).  
	When this ring geometry becomes Coulomb blockaded, and the two leads come into contact with the two wire ends, the current moves through the superconductor or the normal metal as an interferometer. The two-terminal interference conductance can be experimentally measured as a function of gate voltage and magnetic flux through the ring. However, a $4\pi$ periodicity in the conductance-phase relation (often considered the hallmark of MZMs), which can arise both in a topological superconductor and in a trivial metal, cannot establish the existence of MZMs. We show that the trivial metal phase can be ruled out in favor of a topological superconductor by studying persistent conductance distribution patterns.
	In particular, in the presence of MZMs, the conductance peak spacings of the Coulomb blockaded junction would manifest line crossings as the magnetic flux varies. The locations of the line crossings can distinguish line crossings stemming from the trivial metal.


	
\end{abstract}

\date{\rm\today}
\maketitle

Pursuing fault-tolerant quantum computation has been a primary motivation in searching for robust non-Abelian topological excitation in nature. A localized Majorana zero mode (MZM), perhaps the most promising candidate, has been actively studied theoretically~\cite{Sau_semiconductor_heterostructures,Roman_SC_semi,Gil_Majorana_wire,FuKane_SC_STI} and experimentally~\cite{Mourik_zero_bias,RokhinsonLiuFurdyna12,Deng_zero_bias,Churchill_zero_bias,Das_zero_bias,Finck_zero_bias,Albrecht:2016aa}. Although Majorana braiding schemes for quantum computing have been proposed in the literature~\cite{Sarma:2015aa,RMP_braiding,2016arXiv161005289K}, it is unclear at this stage whether sufficient experimental evidence exists providing compelling support for the existence of MZMs. 
In this work, we propose an interference experiment in a Coulomb blockaded topological superconductor ring with magnetic flux in order to provide sufficient and definitive evidence for the existence of MZMs in semiconductor nanowires. It turns out that the theoretical analysis of MZM conductance in a ring structure is quite subtle.

A Coulomb blockade device is small enough so that inserting electrons into the device costs a significant amount of Coulomb charging energy, which dramatically affects the experimental behavior including MZM physics.  
Coulomb blockade physics has been incorporated in the theoretical study of topological superconductors in the context of tunneling transport experiments~\cite{PhysRevB.84.165440,PhysRevB.92.020511,Fu_teleportation,conductance_coulomb_blockade_roman,chiu_blockade}. An important experimental breakthrough in Coulomb blockaded nanowires is the two-terminal conductance measurement in a spin-orbit coupled superconducting semiconductor nanowire 
under an applied magnetic field~\cite{Albrecht:2016aa}. The observed oscillations of the conductance peak spacings~\cite{Albrecht:2016aa} bring new physics to measure MZMs. The superconducting Coulomb blockaded nanowire manifests both $1e$ and $2e$ tunneling signatures through the system with the $2e$ effect presumed to be the ordinary Andreev transport.  
As the lowest energy level of the wire is less than the Coulomb blockade charging energy $E_c$, the $1e$ tunneling dominates transport and exhibits $1e$ periodicity as a function of the gate voltage $V_g$. 
In this manuscript, we focus on the $1e$ tunneling region to study the MZM signatures since this appears to be the prominent transport channel in the Coulomb blockade device.  



Coherent electron \emph{teleportation}~\cite{Teleportatio_Semenoff,Fu_teleportation}, which is the MZM smoking gun signature, has been discussed in the theoretical literature. In this scenario, an electron that is coherently transported from one Andreev Bound state~\cite{PhysRevB.92.020511,2017arXiv170903727D} can be observed by detecting $4\pi$ periodic Aharonov-Bohm(A-B) oscillations, as electrons travel in a ring-like geometry similar to the one shown in Fig.~\ref{Schematic}.  


Here we consider the possibility of the A-B oscillations that would arise because of teleportation. For this purpose, we consider the setup in Fig.~\ref{Schematic}, where an electron transported between the leads L and R can either be teleported through the Majorana wire (shown in blue) or through the normal segment, leading to current interference if the system is phase coherent (i.e.~very low temperature). The entire system is Coulomb blockaded to support a definite number of electrons. When the blue region in Fig.~\ref{Schematic} is an ideal spin-orbit coupled nanowire, the application of a Zeeman field (above a critical field~\cite{Roman_SC_semi,Gil_Majorana_wire,Sau_semiconductor_heterostructures}) should transform the wire into the topological phase with MZMs localized at the wire ends. The transported electrons in the topological phase would have two paths to go from L to R -- one through the wire and the other through the red quantum dot. The interference of the electrons between these distinct paths is expected to manifest itself in an $h/e$ flux dependence of the conductance of the system~\cite{PhysRevB.84.165440,Fu_teleportation,PhysRevB.93.184502,PhysRevB.81.085101}. These oscillations of $4\pi$-periodicity 
can also be thought of as arising from the fractional Josephson effect~\cite{Kwon2004,PhysRevB.92.134516,Kitaev2001,PhysRevB.93.184502,PhysRevB.94.115430} in the ring Josephson junction shown in Fig.~\ref{Schematic}. The Coulomb blockade constrains the number of particles in the system and in some situations may lead to $4\pi$-periodicity of superconducting transport properties of the system~\cite{PhysRevLett.82.3685}.

Specifically, in this paper, we consider a microscopic model for normal transport in the system shown in Fig.~\ref{Schematic}. An important question here is how to distinguish the $4\pi$-periodic MZM oscillations from the trivial A-B oscillations in a normal metal ring. The problem, of course, is that one cannot a priori rule out the possibility of non-superconducting A-B transport in this ring geometry leading to A-B oscillations, which are always $4\pi$-periodic. Despite the fact that the $4\pi$-periodic oscillation (i.e.,~the fractional Josephson effect) strongly distinguishes topological superconductors from conventional superconductors (the regular Josephson effect), $4\pi$ also happens to be the periodicity of A-B oscillations where a non-superconducting metal ring replaces the blue region in Fig.~\ref{Schematic}. A small gap or near critical superconductor has long correlation lengths similar to a normal metal and can also manifest $4\pi$-periodic oscillations. Hence, the observation of the fractional Josephson effect~\cite{WilliamsGoldhaber12,RokhinsonLiuFurdyna12,YamakageTanaka13,Kurter:2015aa} is not conclusive evidence of MZM existence unless normal A-B effect can be decisively ruled out. Solutions proposed to this problem, such as varying the charging energy~\cite{PhysRevB.92.020511}, do not appear to be feasible in the present experimental setups~\cite{Albrecht:2016aa}.
  This motivates us to compare the conductance spectra of such gapless trivial states to the topological system. Interestingly, we find that the conductance of the critical/gapless system indeed manifests $4\pi$-periodic oscillations similar to the topological system. Therefore, $4\pi$-periodic oscillations cannot be the sole means of identifying topological phases. However, a measurement of the excitation gap by comparing the conductance peak spacings, which is a feasible measurement already used in~[\onlinecite{Albrecht:2016aa}], allows one to distinguish the topological phase from the critical/gapless trivial system. \\

The remainder of this paper is organized as follows. In sec.~\ref{setup}, we first establish an interferometer setup for topological superconducting nanowire as well as trivial metal one as a comparison. To propose the observable features of the topological superconductivity and to avoid quasiparticle poisoning, we consider the interferometer is small enough to become Coulomb blockaded and further review the recipe to compute the conductance of the superconducting Coulomb blockade. In sec.~\ref{conductance}, we show the conductance as a function of the magnetic flux for the different interferometers and compare the conductance features of the topological superconductivity with the trivial superconductor and the trivial metal. Sec.~\ref{spacing} is devoted to the study of the conductance peak spacings, which are another observable revealing the topological superconductivity. In sec.~\ref{Vy}, we show the appearance of the MZMs can be manipulated by adjusting the Zeeman field along the spin orbital coupling in the wire. Finally, in Sec.~\ref{conclusion} we conclude the paper and give an outlook on future research.


\section{Interferometer setups}  \label{setup}

\begin{figure}[t!]
\begin{center}
\includegraphics[clip,width=0.59\columnwidth]{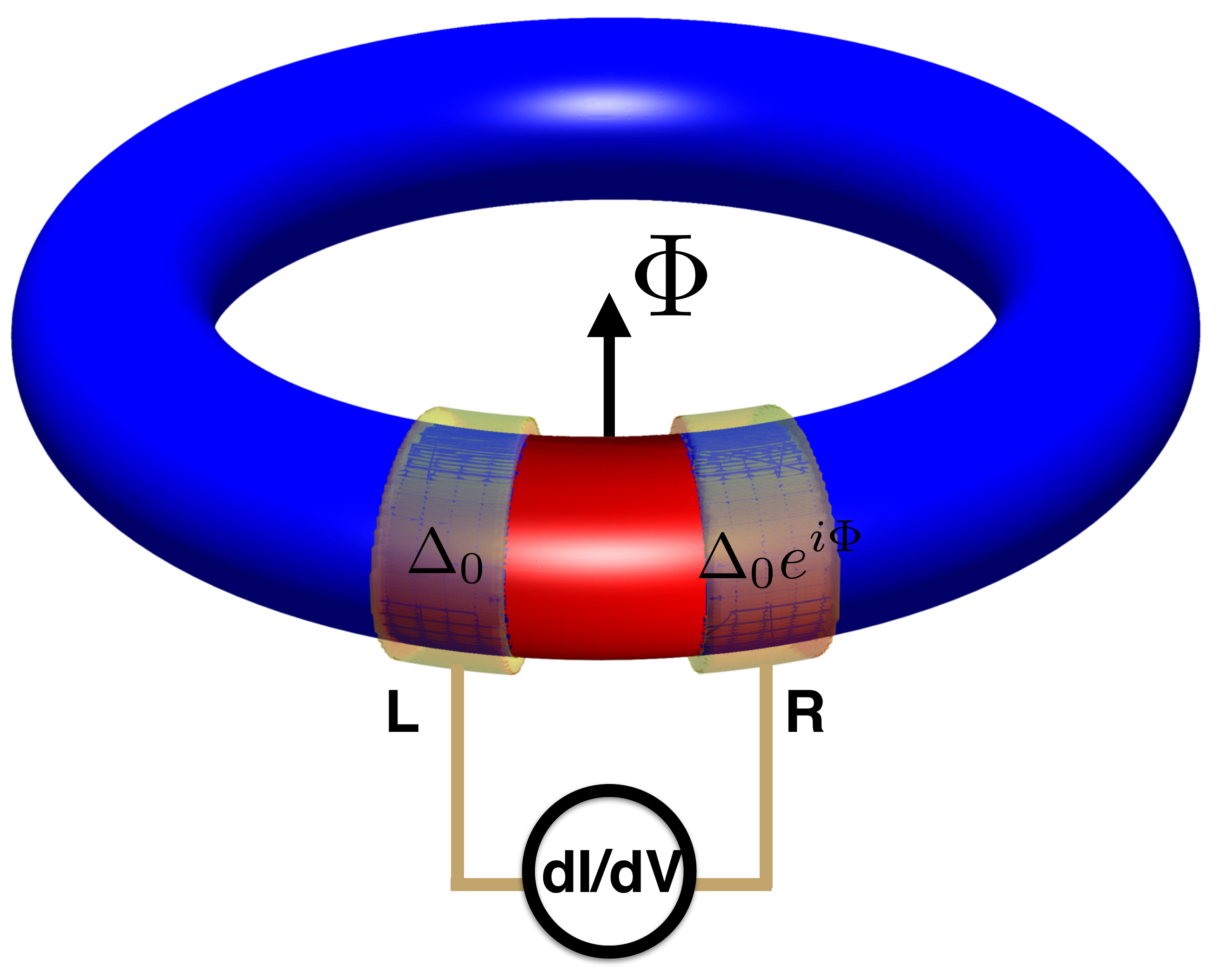}
\end{center}
  \caption{ The schematic of the Coulomb blockaded Majorana ring. The blue region represents the superconducting nanowire ($1\le j \le L$), while the red color ($j=0$) represents a non-superconducting site weakly coupling with the two wire ends. The two leads (dark yellow) come into contact with the wires ends ($j=1, L$) to measure the conductance as the gate voltage $V_g$ applied on this ring and the magnetic flux $\Phi$ goes the middle of the ring. The current moves from one lead to the other through the superconductor or the non-superconducting site as an interferometer. } 
  \label{Schematic}
\end{figure}

The experimental setup we propose is shown schematically in Fig.~\ref{Schematic} with current leads L and R in contact with the ends of a superconducting proximitized nanowire in the presence of spin-orbit coupling. By tuning Zeeman splitting strength, the nanowire passes through the topological quantum phase transition (TQPT), then hosting MZMs at the wire ends. In this topological superconductivity region, we further destroy the MZMs by introducing the coupling between the two wire ends and a non-superconducting metal to form a ring geometry as illustrated in Fig.~\ref{Schematic}, and further insert magnetic flux $\Phi$ in the unit of flux quantum $h/2e$ going through the middle of the ring. The the ring setup is small enough to become Coulomb blockaded.  
In this manuscript, we numerically compute the two-terminal conductance of the Coulomb blockaded ring and propose \emph{observable} and \emph{distinguishable} MZM features.
In particular, we show that the evidence for the topological superconductor hosting MZMs is the line crossings of the conductance peak spacings at the specific values of the superconducting phase differences between the nanowire ends stemming from the inserting magnetic flux. 

 We start with the model of the 1D superconducting proximitized semiconductor nanowire with spin-orbit coupling in the presence of a field-induced Zeeman spin splitting~\cite{Gil_Majorana_wire,Roman_SC_semi}. As its two ends weakly couple with a non-superconducting site, the lattice Hamiltonian can be written as 
\begin{small}
\begin{align}
&\hat{H}_{\rm{BdG}}^{\rm{ring}}= \nonumber  \\ \nonumber 
&\sum_{1\le j \le L} \bigg \{ C^\dagger_j \Big [ \big ( 2t - \mu \big )\tau_z\sigma_0 
+ \Delta_0 \tau_y \sigma_y  +
V_z \tau_z \sigma_z  + V_y \tau_0 \sigma_y \Big ] C_j  \\ \nonumber 
 &+\Big [ C_{j+1}^\dagger  (-t\tau_z\sigma_0 + \alpha i\tau_z \sigma_y )C_{j} +h.c. \Big ] \bigg\} 
\\ &+ w \Big ( C_0^\dagger  \tau_z C_1 +C_0^\dagger  \tau_z C_L+ h.c. \Big )  ,  \label{nanowire Hamiltonian}
\end{align}
\end{small} \noindent
where Pauli matrix $\sigma_\alpha$ represents spin degree of freedom and $C_j=(c_{\uparrow j}, c_{\downarrow j},c_{\uparrow j}^\dagger, c_{\downarrow j}^\dagger)^T$ indicates the vector including the Fermion annihilation and creation operators represented by $\tau_\alpha$. The last term in the Hamiltonian represents a non-superconducting state ($C_0$) weakly coupling with the two ends of the superconducting wire. (We note that this non-superconducting normal arm is about the same length as the superconducting wire. We specially consider large Fermi velocity so that the normal arm is effective short.) The left and right leads come into contact with the first site $C_1$ and the last site $C_L$ for the measurement of the two-terminal conductance as illustrated in Fig.~\ref{Schematic}. The results of the observables closely depend on the value of the Zeeman field ($V_y$) in parallel with the direction of the nanowire spin orbital coupling. To simplify the problem, we consider only the Zeeman field ($V_z$) perpendicular to the direction of the spin orbital coupling first and recover non-zero $V_y$ later. 

\begin{figure}[t!]
\begin{center}
\includegraphics[clip,width=0.95\columnwidth]{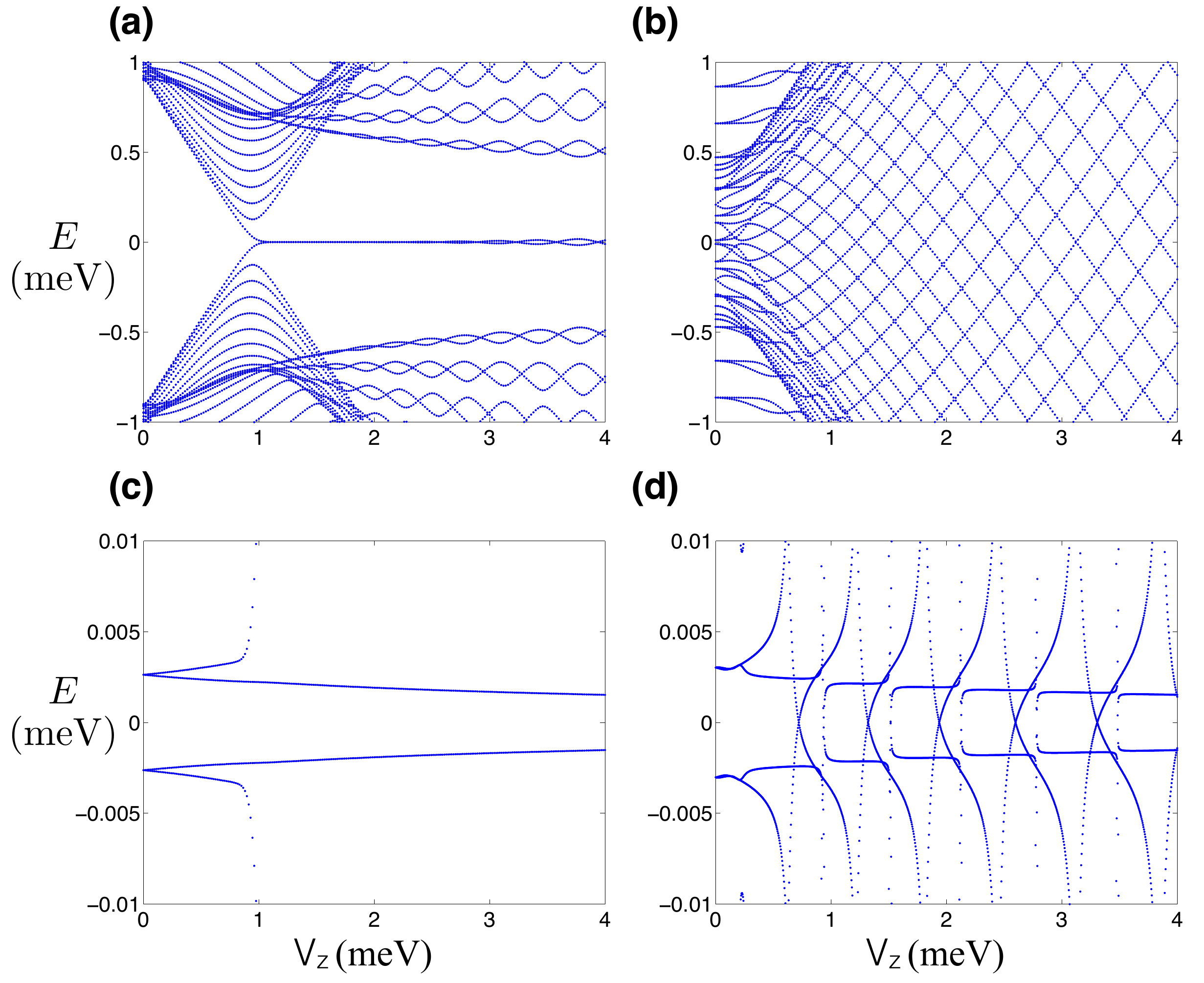}
\end{center}
  \caption{ Energy spectra for the single nanowire and the the interferometer ring as illustrated in Fig.~\ref{Schematic}. In the absence of the non-superconducting site (a,b) represents the energy spectra of the semiconductor nanowire with superconductivity $\Delta_0=0.9$meV and without superconducting gap $\Delta_0 \rightarrow 0$ meV respectively. (c,d) 
  show in the absence of the magnetic flux ($\Phi=0$) the energy spectra of the superconducting and trivial metal rings with the coupling ($w=0.1$meV) between the non-superconducting site ($j=0$) and the superconducting nanowire ends ($j=1,\ L$). This coupling in the superconducting ring always keeps the lowest energy level away from zero, although there is a low-energy state in the non-superconducting site as a normal metal.  
  On the other hand, in non-superconducting ring, as $V_z$ increases, the energy level sometimes reaches to zero energy. 
  The values of the remaining parameters are based on [\onlinecite{chiu_blockade}] with the lattice constant $a=15$nm: hopping strength $t=6$meV (effective mass $=1.5\times 10^4$eV/c$^2$), spin-orbit coupling $\alpha=1.2$meV, superconducting order parameter $\Delta_0=0.9$meV, the chemical potential $\mu=0.2$meV, and the length of the wire $L=80$ ($1.2\mu$m).  
These parameters are used for the following conductance calculation. } 
  \label{spectrum}
\end{figure}

	The magnetic flux $\Phi$ through the middle of the ring can be addressed by the Peierls substitution in the BdG Hamiltonian (\ref{nanowire Hamiltonian}): for $j\neq 0$
\begin{small}
\begin{align}
C^\dagger_j \Delta_0 \tau_y \sigma_y C_j \rightarrow C^\dagger_j  \Delta_0& \Big [ \cos (2j\phi)   \tau_y \sigma_y +  \sin (2j\phi)   \tau_x \sigma_y  \Big ] C_j, \\
C_{j+1}^\dagger  (-t\tau_z\sigma_0 + \alpha i\tau_z \sigma_y )C_{j} &\rightarrow  \nonumber  \\ 
C_{j+1}^\dagger \big[ (\tau_z +\tau_0)e^{-i  \phi}&+ (\tau_z -\tau_0) e^{i  \phi} \big ] \frac{-t\sigma_0 + \alpha i\sigma_y }{2}C_{j},
\end{align}
\end{small} \noindent 
where $\phi=\Phi/2L$. The superconducting phase difference between the two wire ends is given by $\Phi(L-1)/L \sim \Phi$ in large-$L$ limit. By performing the exact diagonalization for $\hat{H}_{\rm{BdG}}^{\rm{ring}}$ of the superconducting nanowire without non-superconducting site ($C_0$), the calculated energy spectrum in Fig.~\ref{spectrum}(a) shows that the TQPT is located at $V_{zc}=\sqrt{\Delta_0^2+\mu^2}=0.922$meV. 
Furthermore, in the presence of the non-superconducting site ($C_0$) and the coupling $w=0.1$meV, the energy level always has a small gap as the Zeeman field ($V_z$) increases as shown in Fig.~\ref{spectrum}(c). 
We label $E_p$ in ascending order as the quasiparticle energy levels of the ring with respect to the BCS ground state and $E_1$ represents the energy difference between the lowest energy BCS states with the odd and even parities. (At $\Phi=0,\ V_z=0$, we choose $E_1>0$ for the BCS ground state with even parity.) 
The Zeeman field $V_z$ and the magnetic flux $\Phi$ vary, the BCS wavefunction adiabatically evolves with its fermion parity remaining fixed. To compare features of the topological superconductivity, we introduce the trivial metal nanowire with superconducting gap $\Delta_0\rightarrow 0$ as the trivial topological phase. The spectra of the metal nanowire and ring exhibits multiple zero energy crossings as shown in Fig.~\ref{spectrum}(b,d).

\begin{figure*}[t!]
\begin{center}
\includegraphics[clip,width=1.65\columnwidth]{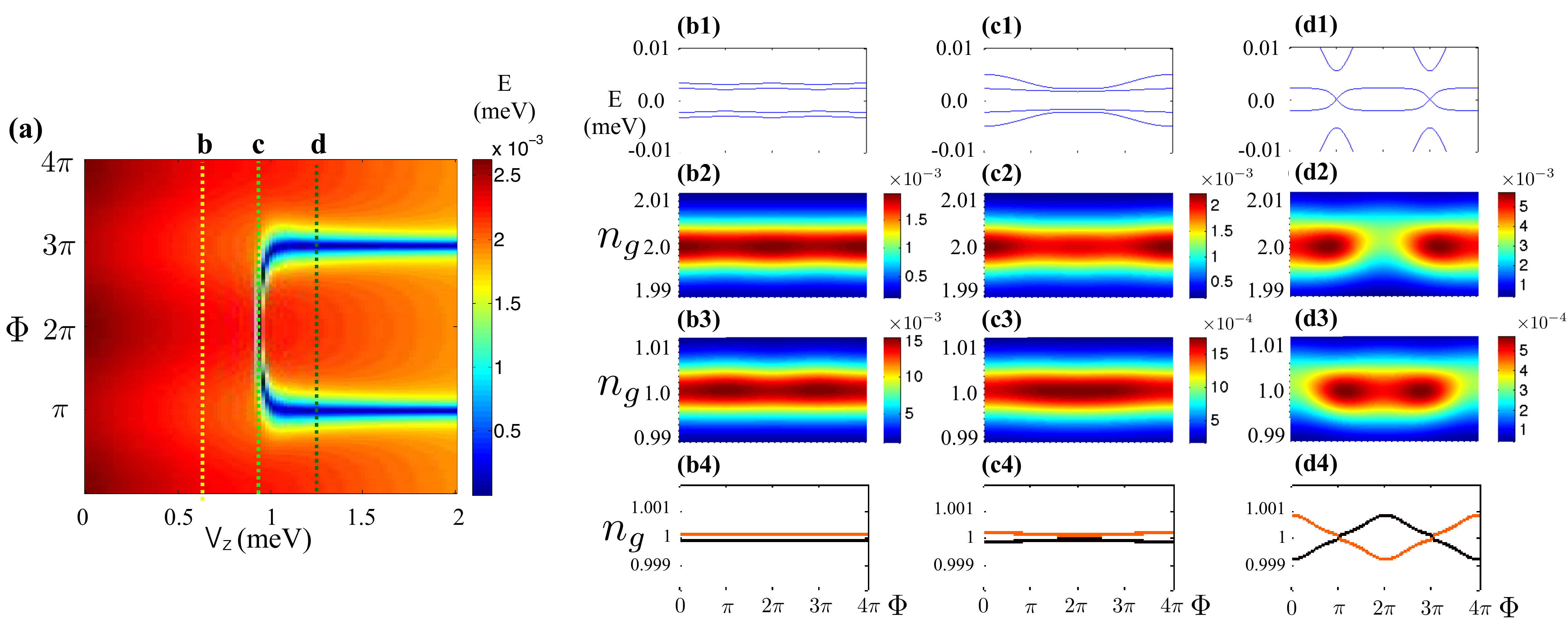}
\end{center}
  \caption{For the superconducting interferometer, quasiparticle and quasihole spectrum, conductance, and conductance peak spacings as the Zeeman potential $V_Z$, which is perpendicular to the spin-orbital direction, is tuned to change the phase from the conventional phase (panels b) ($V_z=0.75$meV), through the TQPT  (panels c)($V_{zc}=0.922$meV) to the topological phase (panels d)($V_z=1.2$meV). Panel (a) shows the positive energy level closest to zero energy and the blue color indicates the presence of the MZMs localized at the ends of the superconducting nanowire at $\Phi=\pi,\ 3\pi$. The two blue color lines merge at the TQPT ($V_{zc}=0.922$meV). Panels (1) represent the energy spectra of the ring at fixed $V_z$'s, while panels (2,3) show the conductance at temperature $T=0.01$meV with even and odd particle number $N$ respectively. As expected all quantities in panels (b1,b2,b3) are $2\pi$ periodic in the flux $\Phi$. In contrast, in the topological phase due to the line crossings ($E_1=0$) at $\Phi=\pi,\ 3\pi$ (panel d1)~\cite{PhysRevB.92.134516,PhysRevB.79.161408},  the conductance oscillations (panels d2,d3), which are $4\pi$ periodic, are distinguishable from the conventional superconductor case. While the spectra are $4\pi$ periodic in the critical region (panel c1), which is similar to a conventional metal, the conductance oscillations (panels c2,c3) are $4\pi$ periodic. Panels (4) represent the conductance peak spacings for even $S_e$ (black) and odd $S_o$ (orange) parities at $T=0.01$meV.  The important MZM feature is that the crossing of the two parity lines are \emph{fixed} at $\Phi=\pi, 3\pi$ as $V_z$ varies in the topological region since panel a shows MZMs are always located at $\Phi=\pi\ ,3\pi$; elsewhere, the lines of the spacings are flat and close to $1$. 
  } 
  \label{transition}
\end{figure*}

Now we consider a situation where the ring becomes Coulomb blockaded with charging energy $E_c$. To compute the conductance, we use the already developed master equation formalism~\cite{chiu_blockade} by assuming that the tunneling rates between the leads and the wire ends are much less than the system temperature ($T$) and the energy level spacing.   
We use the tunneling rates between the leads and the wire ends defined in~[\onlinecite{chiu_blockade}]. Furthermore, we assume that the charging energy $E_c$ ($=2$meV in the following calculation) is large enough so that only the two lowest energy levels $U(N+1)$ and $U(N)$ of the electrostatic energy 
\begin{align}
U(N)=&E_c (N-n_g)^2,  \label{staticenergy}
\end{align} 
are involved in transport with the energy levels of the other electron numbers being too high to be important. 
The gate voltage ($V_g$) of the ring is an experimentally controllable physical parameter, which is proportional to $n_g$ since $n_g=CV_g/e$, where $C$ is the ring capacitance. In the following, the conductance is computed as a function of normalized gate voltage $(n_g)$ and the flux $\Phi$. 
In a superconductor, since the physics is not altered by the transformation $N\rightarrow N+2$ and $n_g \rightarrow n_g+2$ in \cref{staticenergy} (adding a Cooper pair), the conductance as a function of $n_g$ exhibits $2e$ periodicity. We further assume that for the trivial metal ring also has this $2e$-periodicity since the ring is deposited on the top of the superconductor. Hence, computing the conductance with even and odd particle number $(N)$ is enough to describe the interference phenomenon.



We have to define the lowest energy BCS states with $N$ and $N+1$ particle numbers respectively as base states,   
since the Coulomb blockade conductance computed in the master equation formalism~[19] mainly based on these base states. 
At the beginning ($V_z=0$, $\Phi=0$), all of the quasiparticle energy levels $E_i$ are chosen to be positive and the lowest energy BCS state with even particle number is defined to obey 
\begin{equation}
a_{E_p}\ket{\rm{BCS}_e}=0 \label{even BCS}, 
\end{equation}
for all $p$, where $a_{E_p}$ is the quasiparticle annihilation operator. Hence, $\ket{\rm{BCS}_e}$ is the BCS ground state.
 As $V_z$ and $\Phi$ vary, the fermion \emph{parity} of the base state should be fixed and the lowest positive energy level $E_1$ might reach zero as a band crossing.
After this energy band crossing, $\ket{\rm{BCS}_e}$ evolves to the BCS first excited state still obeying \eqref{even BCS} and the value of the energy level $E_1$ is changed to negative from positive. Although the entire qausiparticle and quasihole energy specta ($\pm E_i$) are identical, as the base state with even fermion parity has negative $E_1$, the different quasiparticle energy spectra ($-|E_1|,E_{i\neq 1}$) and ($|E_1|,E_{i\neq 1}$) do lead to the two distinguishable conductances. As $V_z$ and $\Phi$ continuously change, after the next zero energy crossing, the base state with the fixed fermion parity evolves back to the BCS ground state until the third band crossing and so on.

Similarly, at the beginning ($V_z=0$, $\Phi=0$), the lowest energy BCS state with odd particle number obeys 
\begin{equation}
a_{-E_1}\ket{{\rm{BCS}_o}}=0,\  a_{E_{p\neq 1}}\ket{\rm{BCS}_o}=0 \label{odd BCS}, 
\end{equation}
where the quasiparticle annihilation operator $a_{-E}$ with energy $-E$ is equivalent to the qausihole annihilation operator $a_{E}^\dagger$ with energy $E$. After the first gap closes $E_1=0$, with the fixed parity $\ket{\rm{BCS}_o}$ becomes the BCS ground state from the BCS first excited state.  When the second gap closing is passed, $\ket{\rm{BCS}_o}$ goes back to the BCS first excited state. We carry out our numerical calculations following the parity-fixed prescription above.

\section{Coulomb Blockaded Conductance} \label{conductance}

%

\begin{figure*}[t!]
\begin{center}
\includegraphics[clip,width=1.65\columnwidth]{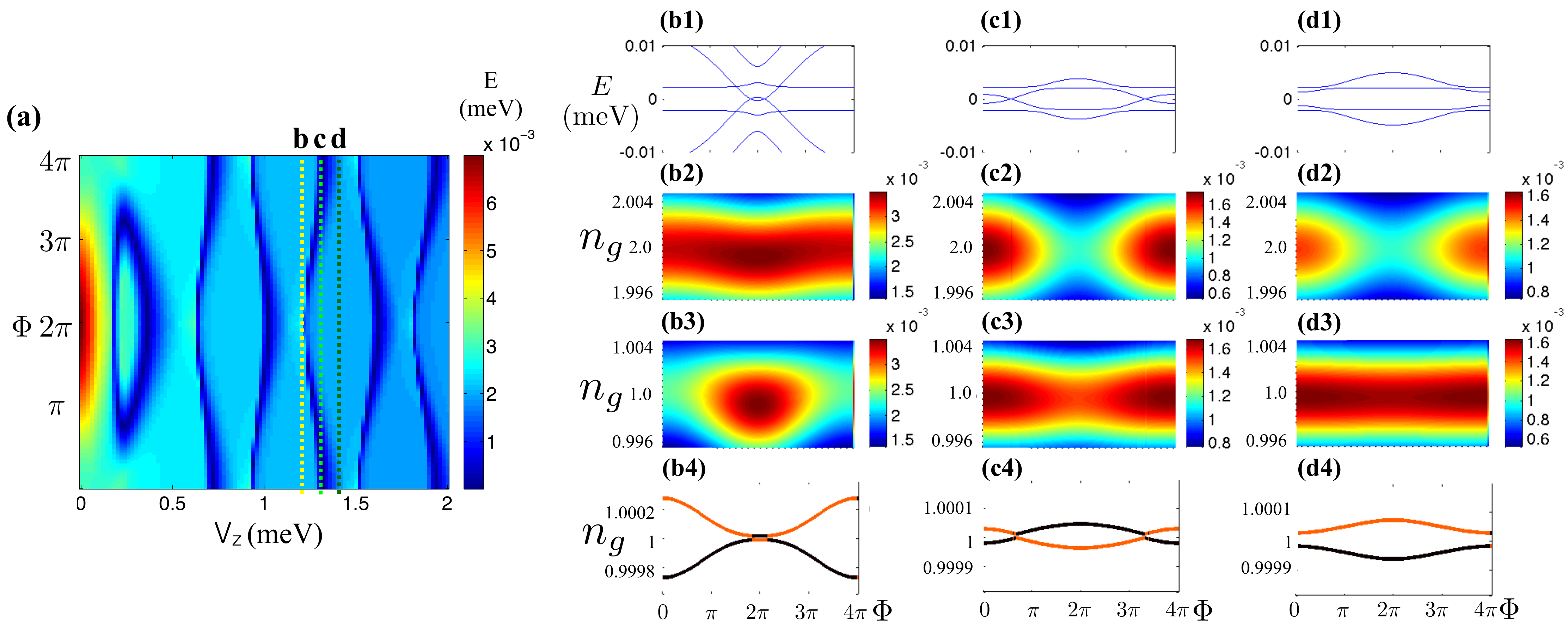}
\end{center}
  \caption{  For the trivial metal interferometer, quasiparticle and quasihole spectrum, conductance, and conductance peak spacings show the features of A-B effect. Panel (a) shows the positive energy level closest to zero energy and the blue color indicates zero energy modes, which are not MZMs. A-B effect exhibits $4\pi$-perodicity of $\Phi$ in the trivial metal interferometer ring $\Delta_0 \rightarrow 0$ for different fixed values of $V_z$: (b) $1.225$meV (c) $1.3$meV (d) $1.375$meV.   Panels (1,2,3,4) represent the identical physical quantities in Fig.~3 respectively. Panels (1) show the $\Phi$-locations of the zero energy modes move around as $V_z$ vary and even the zero energy modes vanish at some $V_z$'s. The zero energy modes can be reflected by the crossing of the conductance peak spacings in panels (4). 
 }
  \label{no_SC}
\end{figure*}

The conductance and the conductance peak spacings can reveal some features of the topological superconducting ring in the trivial ($V_z<V_{zc}$, Fig.~\ref{transition}(b)), TQTP ($V_z=V_{zc}$, Fig.~\ref{transition}(c)), and topological ($V_z>V_{zc}$, Fig.~\ref{transition}(d)) regions. The MZMs with zero energy appear at $\Phi=0,\ 3\pi$ after TQTP as illustrated in Fig.~\ref{transition}(a), which shows the lowest energy level of the superconducting ring as $V_z$ and $\Phi$ vary. 
In the gapped superconductor ($V_z\neq V_{zc}$) the quasiparticle and quasihole energy spectrum ($\pm E_i$) (Fig.\ref{transition}(b1,d1)) always exhibits the $2\pi$-periodicity of the magnetic flux. 
As the flux increases from $0$ to $2\pi$, in the absence of the gap closing (Fig.\ref{transition}(b1)), the quasiparticle energy levels $E_p$ evolve back to the original spectrum. 
Hence, since the spectra with $\Phi=0,\ 2\pi$ are identical, the conductance always has $2\pi$-periodic oscillation. 
The coupling between the normal site and the wire ends completely keeps the entire system gapped ($E_1\neq 0$) in the trivial region ($V_z<V_{zc}$) as shown in the ring spectrum of Fig.~\ref{transition}(a,b1); thus, the conductance oscillation is $2\pi$-periodic and consistent with the numerical result in Fig.~\ref{transition}(b2,b3).


Near the TQTP point, the energy spectrum (Fig.~\ref{transition}(c1)) exhibits the periodicity of $4\pi$. 
Hence, the conductance oscillations in Fig.~\ref{transition}(c2,c3) also are $4\pi$-periodic. This $4\pi$ periodicity is similar with the conductance of the trivial metal ring, which will be discussed later. 




In the topological region ($V_z> V_{zc}$), as the flux is adjusted to $\pi,\ 3\pi$, MZMs appear on the ends (the first and $L$-th sites) of the nanowire~\cite{PhysRevB.79.161408} so that the energy gap is closing ($E_1=0$) as shown in Fig.~\ref{transition}(c1). The lowest quasiparticle energy level $E_1$ is changed to be negative from positive at $\Phi=\pi$ and back to positive as $\Phi$ passes through $3\pi$.  Since the energy level $E_1$ goes back to the original value after $4\pi$ flux, the system with $\Phi=0,\ 4\pi$ is identical. 
Thus, beyond the TQPT point, the $4\pi$-periodic quasiparticle energy spectrum produces $4\pi$-periodic oscillation as shown in in Fig.~\ref{transition}(c2,c3).

To distinguish topological superconductivity from normal-metal A-B effect, we turn off the superconductivity $\Delta_0 \rightarrow 0$ in the nanowire and the lowest energy spectrum of the normal metal ring as a function of $V_z$ and $\Phi$ is shown in Fig.~\ref{no_SC}(a). Due to the nature of the normal metal, the zero energy modes, which are definitely not MZMs, appear several times. The spectrum exhibits the $4\pi$ periodicity of $\Phi$, which can be seen the spectra, as $\Phi$ varies, at the three fixed $V_z$ values in Fig.~\ref{no_SC}. It is expected that the conductances of the Coulomb blockaded normal metal ring have the $4\pi$ periodicity of $\Phi$ in Fig.~\ref{no_SC}(2-3). The normal metal conductance shows the non-superconducting wire has the similar $4\pi$-periodicity patterns with the topological superconductor ($V_z  \geq V_{zc}$) in Fig.~\ref{transition}(c2,c3,d2,d3).

The observation of $4\pi$ periodicity cannot directly lead to the conclusion of the MZM existence, since in the presence of the normal metal 
the conductance oscillation is also trivially $4\pi$-periodic as the manifestation of the usual A-B effect~\cite{PhysRevB.92.020511}. (This superficial agreement between interference phenomena in a topological superconducting ring and a normal metal ring arises simply from both systems manifesting $1e$ coherent transport for different reasons.)  Although the conductance-phase relation has the same $4\pi$ periodicity in both cases, the conductance distributions as a function of $\Phi$ and $n_g$ are distinguishable.
Since in the topological region the lowest energy level (Fig.~\ref{transition}(a)) does not significantly change as $V_z$ varies, the conductance distribution patterns (Fig.~\ref{transition}(d2,d3)) \emph{persist} at any $V_z$ in the entire topological region whereas the patterns alter \emph{dramatically} as $V_z$ varies in the normal metal; these features are confirmed in our numerical simulation. 
The observation of the conductance distribution persistence in the topological region is the key feature distinguishing the topological superconductor and the normal metal. 


\section{Conductance peak spacings}\label{spacing}

The measurement of the conductance peak spacings is an important observable to probe the presence of the MZMs possessing zero energy. The reason to study the conductance peak spacings is that the spacings were successfully measured in the Coulomb blockaded superconducting wire~\cite{Albrecht:2016aa}.	
	The location of the conductance peaks are determined by the maximum conductance values of $n_g$ at fixed flux $\Phi$ and $V_z$.
	At low temperature ($T \ll E_2$), their gate voltage locations for even and odd $N$ are given by $n_g^e(N)=N-E_1/2E_c$ and  $n_g^o(N)=N+E_1/2E_c$ respectively, where the resonant $1e$ tunneling occurs. The important quantities studied in the Coulomb blockade experiment \cite{Albrecht:2016aa} are the even and odd peak spacings
\begin{align}
S_o=&n_g^e(N+1)-n_g^o(N)=1-E_1/E_c, \label{SO} \\
S_e=&n_g^o(N+1)-n_g^e(N)=1+E_1/E_c, \label{SE}
\end{align}
which are the differences between the two closest conductance peaks. The spacings directly depict the lowest energy spectrum. However, when $E_2 \lesssim T$ the 2nd energy level affects the location of the conductance peak, Eqs.~\ref{SO},\ref{SE} do not hold. This is the reason we use the master equation formalism for a superconducting Coulomb blockade to compute conductance in the high-temperature generic situations since the low-temperature constraint is unlikely to be satisfied experimentally~\cite{chiu_blockade}. However, even at higher temperature, when $E_1=0$ in the presence of the MZMs, the even and odd peak spacings~\cite{chiu_blockade} are identical due to the $1e$ tunneling resonance. Hence, the energy level crossings at zero energy in the superconducting spectrum are directly reflected by the crossing of the conductance peak spacings. 



	

\begin{figure}[t!]
\begin{center}
\includegraphics[clip,width=0.99\columnwidth]{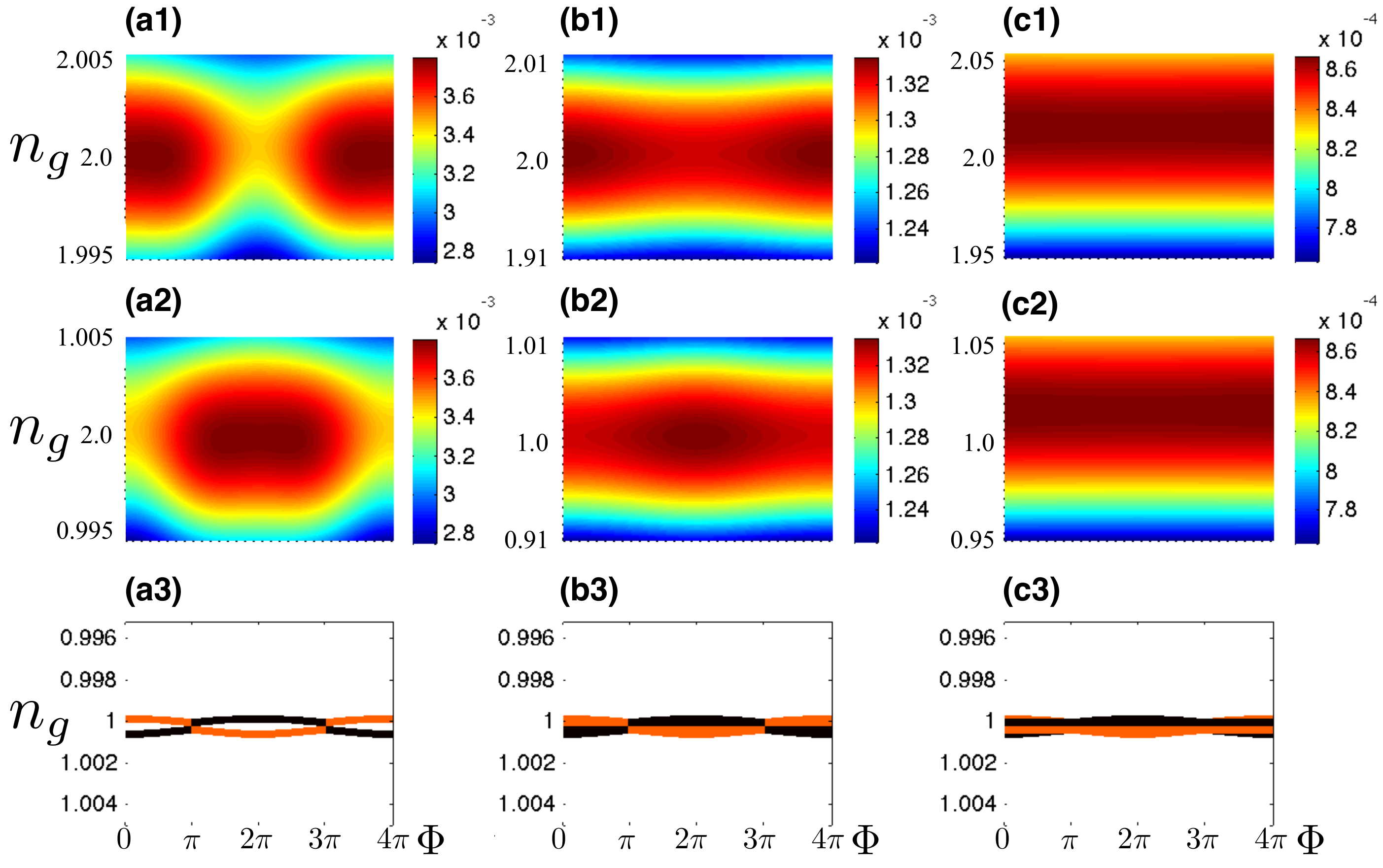}
\end{center}
  \caption{ The conductance and the conductance peak spacing in the topological region $V_z=1.2$meV (Fig.~3(c1)) at different temperatures (a) $T=0.02$meV (b) $T=0.07$meV (c) $T=0.3$meV. Panels (1,2,3) represent the identical physical quantities in Fig.~3(2,3,4) respectively, except for temperature. High temperature broadens the conductance peak and suppresses the oscillation of the conductance peak spacings.   } 
  \label{T_dependence}
\end{figure}



We first discuss the conductance peak spacings for the superconducting ring as shown in Figs.~\ref{transition}(4). In the topological region, the appearance of the MZMs with $E_1=0$ at $\Phi=\pi,\ 3\pi$ leads to the crossing of the two conductance peak spacing lines. It is confirmed by our numerical simulation that this crossing feature can be seen at any $V_z$ value in the topological region. Although,  as shown in Fig.~\ref{transition}(d1) near $\Phi=\pi,\ 3\pi$, Eqs.~\ref{SO},\ref{SE} do not hold due to the second energy level, which is smaller than the temperature ($E_2 < T= 0.01$meV), $4\pi$-periodicity of the conductance peak spacing (Fig.~\ref{transition}(d4)) shares clear similarity with the lowest energy spectrum (Fig.~\ref{transition}(d1)).
On the contrary, in the trivial region and the TQPT, the flatness of the lowest energy level and the suppression of the second lowest energy level ($<T=0.01$meV) lead to flat conductance peak spacing lines without the crossing as shown in Fig.~\ref{transition}(b4,c4), contrasting with the topological situation. 
Therefore, in going from the trivial system to the topological superconductor, the transition of the conductance peak spacing from flat lines to $4\pi$-periodic oscillation and line crossings should be observed.

The normal metal ring also exhibits line crossings in the conductance peak spacings. However, as shown in Figs.~\ref{no_SC}(4), the crossings move to different values of $\Phi$ and even vanish as the Zeeman splitting $V_z$ varies. Since the crossings are not fixed for the trivial metal ring as $V_z$ varies, the fixed line crossings of the conductance peak spacings (at $\Phi=\pi, 3\pi$ only for $V_y=0$) can be the evidence for MZMs.

Ideally, the temperature is expected to be low so that the conductance peak spacings can faithfully depict the lowest energy spectrum. High temperature might alter the observable and disguise the MZM evidence. Fig.~\ref{T_dependence}(1,2) shows that in the topological region, the conductance peak is broadened at higher temperature and the difference between the two conductance peak spacings are suppressed by higher temperature. As the temperature is too high, $4\pi$ periodicity of the conductance-phase relation cannot even be observed as shown in Fig.~\ref{T_dependence}(1,2) and the conductance peak spacing becomes flat as the temperature increases as shown in  Fig.~\ref{T_dependence}(3). 	

\begin{figure*}[t!]
\begin{center}
\includegraphics[clip,width=1.95\columnwidth]{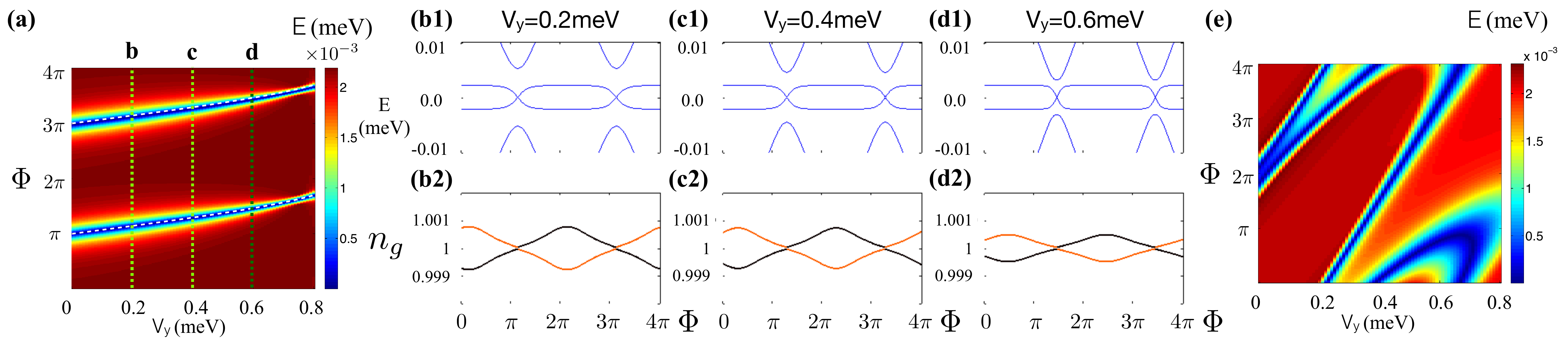}
\end{center}
  \caption{ The physics of the topological superconducting interferometer (a-d) with non-zero Zeeman field $(V_y)$ along the spin orbital coupling is compared with the normal metal interferometer. Panel (a) shows in the topological region at fixed $V_z=1.2$meV the energy level closest to zero energy. The white dashed lines, which are the analytic solution of MZM $\Phi$-location described by eq.\ref{MZM location}, are consistent with the blue color indicating the MZMs. Panels (1) represent the low energy spectra at the different values of $V_y$'s. As $V_y$ increases, the MZMs move away from $\Phi=0,\ 3\pi$. The conductance peak spacings in panels (2) show that the line crossings reflect the presence of the MZMs. On the other hand, the energy level closest to zero energy in the trivial metal interferometer at $V_z=1.225$meV in panel (e) shows that the zero energy modes (blue color) move differently as $V_y$ varies. The $\Phi$-location of the crossing of the conductance peak spacings is an important observable to distinguish the topological superconductivity and the trivial metal. }
  \label{Vy}
\end{figure*}

\section{Zeeman field along the spin orbital coupling direction}\label{Vy}

	For this superconducting semiconductor nanowire interferometer, MZMs with exact zero energy appear at $\Phi=\pi,\ 3\pi$ due to an effective time reversal symmetry. However, in reality this symmetry can be easily broken by applying magnetic field ($V_y$) along the spin-orbital coupling direction. In this section, we derive the flux location of the MZMs as a function of the Zeeman field along the spin-orbital direction. Our derivation scheme is in the following. First, by turning off the coupling ($w$) between the normal site and the ends of the superconducting nanowire, we find the wavefunctions of the MZMs on the nanowire ends in the presence of the magnetic flux. Then we turn on the weak coupling between the two nanowire ends as a first order perturbation. This perturbation energy can faithfully describe the lowest energy in the topological superconducting interferometer.

	We start with the BdG Hamiltonian of the superconducting semiconductor nanowire by Fourier transforming the superconducting part of the Hamiltonian (\ref{nanowire Hamiltonian}) to momentum space  
\bee
H_{\rm{BdG}}(k)=& \big[ 2t(1- \cos k ) - \mu \big] \tau_z \sigma_0 + \Delta_0 \tau_y \sigma_y \nonumber \\
&+ V_z \tau_z \sigma_z + V_y \tau_0 \sigma_y + 2\alpha \sin k \tau_z \sigma_y 
\ee	
We are interested in the low energy theory near the Fermi level as $\mu\approx 2t (1 - \cos k)$. The BdG Hamiltonian can be further simplified to the continuous model 
\bee
H_{\rm{BdG}}(k)\approx 2\alpha k\tau_z \sigma_y  + \Delta_0\tau_y \sigma_y + V_z \tau_z \sigma_z + V_y \tau_0 \sigma_y \quad
\ee
Since the focus is the low energy mode near one nanowire end ($x=L$), the additional phase $\Phi$ of the superconducting order parameter stemming from the magnetic flux can be assumed to be a constant. With this additional phase, the order parameter is given by $\Delta_0 \rightarrow \Delta_0 e^{i\Phi}$. The low energy Hamiltonian can be written as 
\begin{align}
H_{\rm{BdG}}(x) =& 2\alpha\tau_z \sigma_y  \frac{\partial}{i\partial x} +\Delta_0 \cos \Phi \tau_y \sigma_y  \nonumber \\
&+ \Delta_0 \sin \Phi \tau_x \sigma_y  + V_z \tau_z \sigma_z + V_y \tau_0 \sigma_y.
\end{align}
The two wire ends are located at $x=0,\ L$ as domain walls by assuming $ \sqrt{\Delta^2_0- V_y^2}-V_z >0 $ as $0<x<L$ and $ \sqrt{\Delta^2_0- V_y^2}-V_z <0 $ elsewhere. By solving the eigenvalue problem at zero energy, this assumption leads to an unnormalized MZM wavefunction localized at $x=0$ without the additional superconducting phase ($\Phi=0$)
\bee
\ket{\phi_0}=e^{-\frac{R-V_z}{2\alpha}x}
\bma 
-i e^{i\beta/2} \\
-i e^{i\beta/2} \\
i e^{-i\beta/2}  \\
i e^{-i\beta/2} \\
\ema,
\ee
where $R=\sqrt{\Delta^2_0- V_y^2}$ and $\beta=\arctan \frac{V_y}{R}$. In addition, another MZM wavefunction localized at $x=L$ is given by 
\bee
\ket{\phi_L}=e^{\frac{R-V_z}{2\alpha}(x-L)}
\bma 
 i e^{-i\beta/2+i\Phi/2} \\
-i e^{-i\beta/2+i\Phi/2} \\
i e^{i\beta/2-i\Phi/2}  \\
-i e^{i\beta/2-i\Phi/2} \\
\ema
\ee
Since the coupling between the two wire ends is off, as long as the wire length $L$ is long enough, the energies of the two MZMs are close to zero. We turn on the weak coupling between the two ends as the extension of spin orbital coupling 
\bee
\Delta \hat{h}= i\delta ( C^\dagger_0 \tau_z \sigma_y C_L -C^\dagger_L \tau_z \sigma_y C_0 )
\ee  
Consider this term as the first order perturbation, the low energy effective Hamiltonian can be written as the coupling sandwiched by the two MZMs 
\begin{align}
\Delta H =& \bma 
\bra{\phi_0} \Delta \hat{h} \ket{\phi_0} & \bra{\phi_0} \Delta \hat{h} \ket{\phi_L} \\
 \bra{\phi_L} \Delta \hat{h} \ket{\phi_0} & \bra{\phi_L} \Delta \hat{h} \ket{\phi_L}
\ema \\
\propto & 4\delta \sin (\beta -\Phi/2)
\bma
0 & -i \\
i & 0 \\
\ema
\end{align}
The energy level crossing (zero energy) occurs as $\Delta H=0$ at 
\bee
\Phi=(3)\pi+2\beta. \label{MZM location}
\ee
 Hence, as $V_y=0$, the MZMs appear at $\Phi=\pi,\ 3\pi$; otherwise, the MZMs are not fixed at $\Phi=\pi,\ 3\pi$. 

Back to the simulation of the superconducting ring, we recover the Zeeman field ($V_y$) along the spin orbital coupling direction and adjust $V_z$ to the topological region. The interferometer ring is still described by the Hamiltonian (\ref{nanowire Hamiltonian}) with non-zero $V_z$. The energy spectra in Fig.~\ref{Vy}(a,1) show that the MZMs move away from $\Phi= \pi,\ 3\pi$ and the locations of the MZMs are consistent with the analytic solution (\ref{MZM location}). Hence, this analytic solution also predicts to the crossing locations of the conductance peak spacings as shown in Fig.~\ref{Vy}(2). That is, once the physical values of $V_y$ and $\Delta_0$, which determine $\beta$, are known, if $\Phi=(3)\pi+2\beta$ are identical to the location of the crossings observed, this directly supports the existence of the MZMs. We can further compare the MZM locations with the lowest energy spectrum of the trivial metal ring as shown in Fig.~\ref{Vy}(e). The zero energy modes, which are not the MZMs, lead to the crossings of the conductance peak spacings and the distribution of these zero modes is completely different from the topological superconductor hosting MZMs. Therefore, the locations of the observed crossings can clearly distinguish the topological superconducting interferometer from the trival metal interferometer.

\section{conclusion}	\label{conclusion}
	As the Zeeman field $V_z$ increases, observing the transition of the phase-current relation from $2\pi$ periodicity to $4\pi$ periodicity is the preliminary step. However, the transition from a superconductor to a normal metal shares the same periodicity. The details of the observable conductance can exclude the trivial metal scenario. The conductance distribution patterns in the topological region do not alter as a function of $V_z$, whereas the conductance patterns of the normal metal always change as a function of $V_z$. 
	In the topological region the line crossing of the conductance peak spacings, which reflect the $\Phi$-location of the MZM, always occurs at $\Phi=\pi,\ 3\pi$ in the absence of the Zeeman field along the spin orbital coupling direction, while the crossing locations of the trivial metal change with $V_z$. Even in the presence of the Zeeman field along the spin orbital coupling direction, the $\Phi$-location of the line crossing can be analytically predicted. 
	These conductance interference features can serve to distinguish the topological nanowire ring from a normal metal experimentally.



	

This work is supported by Microsoft and Laboratory for Physical Sciences.

\bibliographystyle{apsrev4-1}
\bibliography{TOPO}

\end{document}